\documentclass[12pt]{article}
\usepackage{graphicx}
\usepackage{amssymb}
\newlength{\extraspace}
\newlength{\extraspaces}
\newcommand{\be}{\begin{equation}
\addtolength{\abovedisplayskip}{\extraspaces}
\addtolength{\belowdisplayskip}{\extraspaces}
\addtolength{\abovedisplayshortskip}{\extraspace}
\addtolength{\belowdisplayshortskip}{\extraspace}}
\newcommand{\ee}{\end{equation}}

\newcommand{\ba}{\begin{eqnarray}
\addtolength{\abovedisplayskip}{\extraspaces}
\addtolength{\belowdisplayskip}{\extraspaces}
\addtolength{\abovedisplayshortskip}{\extraspace}
\addtolength{\belowdisplayshortskip}{\extraspace}}
\newcommand{\ea}{\end{eqnarray}}

\newcommand{\newsection}[1]{
\vspace{12mm}
\pagebreak[3]
\addtocounter{section}{1}
\setcounter{equation}{0}
\setcounter{subsection}{0}
\noindent{\bf \thesection. #1}
\nopagebreak
\medskip
\nopagebreak}

\newcounter{saveeqn}

\newcommand{\dif}{\mathrm{d}}
\newcommand{\me}{\mathrm{e}}

\begin{document}
\addtolength{\baselineskip}{1.5mm}

\thispagestyle{empty}
\begin{flushright}

\end{flushright}
\vbox{}
\vspace{2cm}

\begin{center}
{\LARGE{Black holes on gravitational instantons%\\[2mm]
        }}\\[16mm]
{Yu Chen~~and~~Edward Teo}
\\[6mm]
{\it Department of Physics,
National University of Singapore, %\\[1mm]
Singapore 119260}\\[15mm]

\end{center}
\vspace{2cm}

\centerline{\bf Abstract}
\bigskip
\noindent
In this paper, we classify and construct five-dimensional black holes on gravitational instantons in vacuum Einstein gravity, with $\mathbb{R}\times U(1)\times U(1)$ isometry. These black holes have spatial backgrounds which are Ricci-flat gravitational instantons with $U(1)\times U(1)$ isometry, and are completely regular space-times outside the event horizon. Most of the known exact five-dimensional vacuum black-hole solutions can be classified within this scheme. Amongst the new space-times presented are static black holes on the Euclidean Kerr and Taub-bolt instantons. We also present a rotating black hole on the Eguchi--Hanson instanton.

%\addtolength{\baselineskip}{3mm}   %double-spacing

\newpage

\newsection{Introduction}

Unlike their four-dimensional counterparts, black holes in five and higher dimensions are known to exhibit much richer behaviour. For example, they are no longer subject to stringent uniqueness theorems, and can in fact have non-spherical event-horizon topology. The first example of this was the five-dimensional black ring discovered by Emparan and Reall in 2002 \cite{Emparan:2001}, which has an $S^1\times S^2$ horizon topology and can in certain cases carry the same mass and angular momentum as the spherical Myers--Perry black hole \cite{Myers:1986}. Since this discovery, there has been much progress in the study of higher-dimensional black holes and their possible phases (see, e.g., \cite{Emparan:2008,Obers:2008,Rodriguez:2010} and references therein).

The vast majority of the known black-hole solutions to date, in both four and higher dimensions, are asymptotically flat in the sense that they approach a Minkowski space-time $M^{1,n}$ at infinity. This is physically expected of an isolated gravitating system. And indeed, when the black hole is removed from the space-time (for example, by setting its mass to zero), the resulting background space-time is nothing but a direct product of $n$-dimensional Euclidean space $E^n$ and a flat time dimension.

In higher dimensions, however, black holes can admit a variety of other asymptotic behaviour, while still becoming ``flat'' at infinity. In five dimensions, for example, it is possible to consider black-hole solutions that are asymptotically $M^{1,3}\times S^1$, or even those that are asymptotically a finite but non-trivial $S^1$ fibre bundle over $M^{1,3}$. Such solutions are of interest in Kaluza--Klein theory where the fifth dimension is assumed to be compactified into a circle. Examples of such solutions are so-called black holes and black rings on Taub-NUT space, and they have been well-studied in the literature (see, e.g., \cite{Elvang:2005,Gaiotto:2005,Bena:2005,Ishihara:2005,Wang:2006}). When the black hole/ring is removed, the resulting background space-time is a direct product of the (Euclidean) self-dual Taub-NUT space \cite{Newman:1963,Hawking:1976} and a flat time dimension.

In general, the background of a five-dimensional vacuum black-hole space-time can be the direct product of any regular Ricci-flat four-manifold and a flat time dimension. Such four-manifolds are known as {\it gravitational instantons\/}, of which flat space and the self-dual Taub-NUT space are but the simplest examples. Other well-known examples of gravitational instantons include the Euclidean Schwarzschild and Kerr instantons \cite{Gibbons:1976}, the Eguchi--Hanson instanton \cite{Eguchi:1978} and the Taub-bolt instanton \cite{Page:1979}. Conversely, given any (Ricci-flat) gravitational instanton, we can add a flat time dimension to it to obtain a five-dimensional space-time that solves the vacuum Einstein equations.
Moreover, static or stationary black holes may be added into such a space-time. The black holes are then said to be sitting on that corresponding gravitational instanton.

The study of gravitational instantons in the late 1970s has given us a way to describe their possible asymptotic structures \cite{Gibbons:1979gd} (see also \cite{Chen:2010}). There are four known possibilities: asymptotically flat (AF), asymptotically locally flat (ALF), asymptotically Euclidean (AE) and asymptotically locally Euclidean (ALE). The asymptotic behaviour of the gravitational instanton is directly related to the asymptotic behaviour of a black hole on this gravitational instanton. Black holes on the unique AE gravitational instanton $E^4$ will be asymptotically flat in the usual sense, whereas black holes on ALE gravitational instantons (such as Eguchi--Hanson) will be asymptotically locally flat, i.e., they approach $M^{1,4}/\mathbb{Z}_p$ for some integer $p\geq2$ at infinity. Black holes on AF gravitational instantons (such as Euclidean Schwarzschild and Kerr) or ALF gravitational instantons (such as self-dual Taub-NUT and Taub-bolt) will have a space-like direction with finite norm at infinity.

So far, there has only been limited success in constructing five-dimensional black holes on gravitational instantons other than flat space and the self-dual Taub-NUT space. One of the few known ways to systematically construct black holes on a non-trivial gravitational instanton is within some five-dimensional supergravity theory, say $N=1$ minimal supergravity. In such a theory, an underlying linear structure allows the superposition of supersymmetric black holes on any four-dimensional hyper-K\"ahler manifold \cite{Gauntlett:2002}, which includes the self-dual Taub-NUT space and the Eguchi--Hanson instanton. This was the method used to construct black holes/rings on the latter gravitational instanton in \cite{Tomizawa:2007a,Tomizawa:2008tj}. In a similar fashion, extremal black holes on the Eguchi--Hanson instanton \cite{Ishihara:2006} and the multi-Taub-NUT instanton \cite{Ishihara:2006b} have been constructed within the context of five-dimensional Einstein--Maxwell theory.

In this paper, we are interested in five-dimensional black holes on gravitational instantons in vacuum Einstein gravity, with non-degenerate event horizons. The construction of such solutions presents more difficulty than in the supersymmetric or Einstein--Maxwell case, although progress is still possible if a $\mathbb{R}\times U(1)\times U(1)$ isometry is assumed. In this case, the inverse-scattering method \cite{Belinski:2001,Pomeransky:2005} has proven to be a powerful way to generate new exact solutions, by applying soliton transformations on some seed solution. The seed solution in general can be chosen to have a diagonal metric, and can be constructed using the generalised Weyl formalism developed in \cite{Emparan:2001b}. To analyse the solutions thus generated, which have $\mathbb{R}\times U(1)\times U(1)$ isometry, the rod-structure formalism developed in \cite{Emparan:2001b,Harmark:2004,Hollands:2007,Hollands:2008,Chen:2010} provides us a very useful tool. In particular, in our previous paper \cite{Chen:2010}, a stronger version of the rod structure---in which the rod directions are appropriately normalised---was introduced, and we showed how the regularity conditions of these solutions can be read off from it. In this paper we will use this formalism extensively, so it may be worthwhile to give a brief review of it here. The reader is referred to \cite{Chen:2010} and references therein for more details.

In the solutions mentioned above, there exists three linearly independent and mutually commuting Killing vector fields, say $V_{(i)}$ for $i=0,1,2$, which generate the $\mathbb{R}\times U(1)\times U(1)$ isometry of the space-times. We further assume that $V_{(0)}=\frac{\partial}{\partial t}$ generates the time translation symmetry $\mathbb{R}$ and is normalised at infinity. Under suitable conditions, these solutions admit the following metric form in so-called Weyl--Papapetrou coordinates
\be
\dif s^2=G_{ij}\dif x^i\dif x^j+\me^{2\nu}(\dif\rho^2+\dif z^2)\,,
\ee
such that $V_{(i)}=\frac{\partial}{\partial x^i}$, and subject to the constraint $\rho=\sqrt{|\det G|}$. Here $G_{ij}$ and $\nu$ are functions of $\rho$ and $z$ only. It can be seen that the matrix $G(\rho,z)$ is non-degenerate as long as $\rho>0$.  At $\rho=0$, it becomes degenerate, so the kernel of $G(\rho=0,z)$ becomes non-trivial, i.e., ${\rm dim}({\rm ker}(G(0,z)))\geq 1$. It can be shown that in order to avoid curvature singularities, it is necessary that ${\rm dim}({\rm ker}(G(0,z)))=1$, except for isolated values of $z$. When this applies, we label these isolated values as $z_1,z_2,\dots,z_N$, with $z_1<z_2<\cdots<z_N$, and call the corresponding points on the $z$-axis ($\rho=0,z=z_i$) {\it turning points\/}. These turning points divide the $z$-axis into $N+1$ intervals $(-\infty,z_1]$, $[z_1,z_2]$,\dots, $[z_{N-1},z_N]$, $[z_N,\infty)$. These intervals are known as {\it rods\/}, and are labelled from left to right as rod 1, rod 2, \dots, rod $N+1$.

In the interior of a specific rod for $(\rho=0,z_k<z<z_{k+1})$, we can find in the kernel of $G(\rho=0,z)$ a constant vector field $v=v^i\frac{\partial}{\partial x^i}$, such that $G(0,z)v=0$, and such that the (Euclidean) surface gravity on this rod,
\be
\kappa_{(E)}=\lim\limits_{\rho\rightarrow 0}\sqrt{\left|\frac{G_{ij}v^i v^j}{\rho^2\me^{2\nu}}\right|}\,,
\ee
is normalised to unity.
The vector field $v$ is assigned to this specific rod and is called its (normalised) direction. The rod is said to be time-like or space-like, if its direction $v$ generates time-like or space-like flows respectively in the vicinity of this rod. For a given solution, the specification of the rods and their directions is referred to as the {\it rod structure\/} of the solution. Notice that we have the freedom to choose the three linearly independent and mutually commuting Killing vector fields as $(\tilde{V}_{(0)}=V_{(0)},\tilde{V}_{(1)},\tilde{V}_{(2)})$, which are linear combinations of the previous ones, so that we can define the corresponding Weyl--Papapetrou coordinates $(\tilde{x}^i,\tilde{\rho},\tilde{z})$ such that $\tilde{V}_{(i)}=\frac{\partial}{\partial \tilde{x}^i}$. In this way, we can introduce the rod structure {\it in standard orientation\/} by making some particular choices of the Killing vector fields $\tilde{V}_{(i)}$, in which specific rod directions take simple forms according to the rules prescribed in \cite{Chen:2010}.

A time-like rod physically represents a Killing horizon and a space-like rod represents an axis for its direction. To avoid possible conical, orbifold and Dirac--Misner singularities in the space-time, firstly, the orbits generated by the direction of a space-like rod must be identified with period $2\pi$. Secondly, direction-pairs of adjacent space-like rods must be related by $GL(2,\mathbb{Z})$ transformations, i.e.,
\be
(v_k,v_{k+1})=GL(2,\mathbb{Z})(v_l,v_{l+1})\,,
\ee
if the $k$-th, $(k+1)$-th, $l$-th and $(l+1)$-th rods are all space-like. The direction-pair of any two adjacent space-like rods, say $(v_k,v_{k+1})$, can then be identified as the pair of independent $2\pi$-periodic generators of the $U(1)\times U(1)$ isometry group of the space-time. Thirdly, any other space-like rod must have a direction that can be expressed as $av_k+bv_{k+1}$ for some coprime integers $a$ and $b$. This will ensure that there exists a basis in which the direction of every space-like rod will have a vanishing time component. If this were not the case, Dirac--Misner singularities will be present, and certain identifications will have to be imposed on the time coordinate \cite{Misner:1963}.

We finally recall that the rod structure in fact characterises the boundary $\rho=0$ of the orbit space $M/(\mathbb{R}\times U(1)\times U(1))$ of the space-time $M$, which is homeomorphic to the upper-half complex plane parameterised by the coordinates $(\rho,z)$. It encodes much useful information about the solution. In particular, the space-time, as a manifold with $\mathbb{R}\times U(1)\times U(1)$ action, is uniquely determined by the rod structure and can in principle be reconstructed from it.

Happily, it turns out that most of the known gravitational instantons have a $U(1)\times U(1)$ isometry. In \cite{Chen:2010}, we applied the rod-structure formalism to the study and classification of such gravitational instantons. Static or stationary black holes on these gravitational instantons will then have a $\mathbb{R}\times U(1)\times U(1)$ isometry, and can be analysed using the rod-structure formalism. In this paper, we will classify and construct these black-hole solutions. As we will see, most of the known exact five-dimensional vacuum black holes in the literature can be classified within this scheme.

In the following sections, we will classify/construct black holes on four-dimensional flat space, the self-dual Taub-NUT, Euclidean Schwarzschild, Euclidean Kerr, Eguchi--Hanson and Taub-bolt instantons, respectively. The static black holes on the Euclidean Kerr and Taub-bolt instantons, and the rotating black hole on the Eguchi--Hanson instanton constructed in this paper are essentially new five-dimensional space-times. Although their local metrics have previously appeared in the literature, they had either not been studied at all, or interpreted as singular space-times containing conical singularites. We emphasise that the space-times constructed here have the necessary coordinate identifications to ensure that they are completely regular outside the black-hole event horizon. 

For each case, we will analyse the rod structure of the solution. Since the metrics considered in this paper are independent of the three coordinates $(t,\psi,\phi)$, we naturally take the three linearly independent and mutually commuting Killing vector fields to be $\{V_{(0)}=\frac{\partial}{\partial t},V_{(1)}=\frac{\partial}{\partial \psi},V_{(2)}=\frac{\partial}{\partial \phi}\}$. The corresponding Weyl--Papapetrou coordinates are defined as $(x^0=t,x^1=\psi,x^2=\phi,\rho,z)$. The direction of a rod then has the form $a_0 \frac{\partial}{\partial t}+a_1 \frac{\partial}{\partial \psi}+a_2 \frac{\partial}{\partial \phi}$, which will be written as $(a_0,a_1,a_2)$ for simplicity. Where necessary, new Weyl--Papapetrou coordinates $(\tilde{x}^0=\tilde{t},\tilde{x}^1=\tilde{\psi},\tilde{x}^2=\tilde{\phi},\tilde{\rho},\tilde{z})$ are also introduced to put the rod structures in standard orientation.

\newsection{Black holes on four-dimensional flat space}

Five-dimensional Minkowski space-time is obtained by adding a flat time dimension to four-dimensional flat space. The simplest black hole sitting on four-dimensional flat space is then the five-dimensional Schwarzschild black hole \cite{Tangherlini:1963}. In spherical polar coordinates it has the following form:
\ba
\label{bh_4D_Euclid}
{\dif s}^{2}&=&- \left( 1-{\frac {2 m}{{r}^{2}}} \right) {{\dif t}}^{2}+\left( 1-{\frac {2 m}{{r}^{2}}} \right) ^{-1}{{\dif r}}^{2}+{r}^{
2} \left( {{\dif \theta}}^{2}+ \sin ^{2} \theta \,
 {{\dif \psi}}^{2}+\cos^{2} \theta \, {{\dif \phi}}^{2} \right),
\ea
where the parameter $m$ and coordinates $t$, $r$, $\theta$ take the ranges $m\geq 0$, $-\infty<t<\infty$, $r\geq r_0\equiv\sqrt{2 m}$, $0\leq \theta \leq \frac{\pi}{2}$. The black-hole horizon and physical infinity are located at $r=r_0$ and $r=\infty$, respectively.

The Weyl--Papapetrou coordinates $(t,\psi,\phi,\rho,z)$ are related to the above coordinates by
\be
\label{Weyl--Papapetrou_bh_4D_Euclid}
\rho=\frac{1}{2}\,{r}^{2}\sqrt{1-\frac{2 m}{r^2}}\sin 2\theta \,,\qquad z=\frac{1}{2}\,{r}^{2} \left(1-\frac{m}{r^2}\right) \cos 2\theta \,.
\ee
In these coordinates, the rod structure has two turning points, at $(\rho=0,z=z_1\equiv-\frac{m}{2})$ or $(r=r_0,\theta=\frac{\pi}{2})$, and $(\rho=0,z=z_2\equiv\frac{m}{2})$ or $(r=r_0, \theta=0)$. These two turning points correspond to the south and north poles of the black-hole horizon, respectively. From left to right, the three rods are:

\begin{itemize}
\item Rod 1: a semi-infinite space-like rod located at $(\rho=0, z\leq z_1)$ or $(r\geq r_0, \theta=\frac{\pi}{2})$, with direction $\ell_1=(0,0,1)$.

\item Rod 2: a finite time-like rod located at $(\rho=0, z_1\leq z\leq z_2)$ or $(r=r_0, 0\leq \theta \leq \frac{\pi}{2})$, with direction $\ell_2=\frac{1}{\kappa}(1,0,0)$, where $\kappa=\frac{1}{\sqrt{2m}}$.

\item Rod 3: a semi-infinite space-like rod located at $(\rho=0, z\geq z_2)$ or $(r\geq r_0,\theta=0)$, with direction $\ell_3=(0,1,0)$.
\end{itemize}
This rod structure is illustrated in Fig.~1 (with a tilde added to the $z$ coordinate). Since the second rod is time-like, it represents the black-hole horizon. The two semi-infinite rods 1 and 3 are space-like, and represent the two asymptotic axes.

We can identify the orbits generated by $\{\ell_1, \ell_3\}$ with period $2\pi$ independently to ensure regularity of the space-time. However, it is interesting to note that since the two space-like rods do not intersect at a turning point, we can further identify the orbits generated by a third Killing vector field $\frac{1}{s}(q\ell_1+p\ell_3)$ with period $2\pi$ without introducing either conical or orbifold singularities into the space-time, provided that $s$, $p$ and $q$ are mutually coprime non-zero integers. The resulting space-time will have an asymptotic structure that has the same topology as that of the black-hole event horizon, namely a lens-space $L(s,nq)$ where the integer $n$ solves $np=1$(mod $s$). By removing the black hole from this space-time (by setting $m=0$), we recover a quotiented five-dimensional Minkowski space-time background $M^{1,4}/\mathbb{Z}_{|s|}$, which {\it is\/} singular if $|s|\geq 2$, as there will be a $\mathbb{Z}_{|s|}$ orbifold singularity present at the origin.

In this paper, the orbits generated by $\{\ell_1, \ell_3\}$ are identified with period $2\pi$ independently, so the following identifications:
\be
\label{identification for 4D Euclid}
(\psi, \phi) \rightarrow (\psi, \phi+2 \pi) \,,\qquad (\psi, \phi) \rightarrow (\psi+2 \pi, \phi) \,,
\ee
are made to ensure the regularity of the black-hole space-time (\ref{bh_4D_Euclid}), as well as its background. The resulting space-time is just the five-dimensional Schwarzschild black hole, whose $U(1)\times U(1)$ isometry group is generated by the two independent $2\pi$-periodic generators $\{\ell_1, \ell_3\}$. It is then obvious that this black hole has horizon-topology $S^3$, and is asymptotically flat.

Angular momenta can be added to the black hole along the two orthogonal axes at infinity, giving the five-dimensional Myers--Perry black hole \cite{Myers:1986}. In this case, the direction of the time-like rod is rotated, i.e., it has components involving $\ell_1$ and $\ell_3$, so the black hole has non-vanishing angular velocities along the two space-like directions.

The five-dimensional Schwarzschild and Myers--Perry black holes sit at the turning point of four-dimensional flat space (thus covering it). A black hole can also sit on, say, the left semi-infinite rod of four-dimensional flat space. In terms of the rod structure, this is done by cutting the left semi-infinite rod, and placing a time-like rod there. Angular momenta can also be added to the black hole to balance its self-gravitation. The resulting solutions are the single-rotating Emparan--Reall \cite{Emparan:2001} and double-rotating Pomeransky--Sen'kov \cite{Pomeransky:2006} black rings. We can also perform a series of these operations, to get the so-called black saturn \cite{Elvang:2007}, black di-ring \cite{Iguchi:2007,Evslin:2007} and black bi-ring \cite{Izumi:2007,Elvang:2007b} solutions.  By removing the black holes/rings from these configurations, we recover five-dimensional Minkowski space-time.

\begin{figure}[t]
\begin{center}
\includegraphics{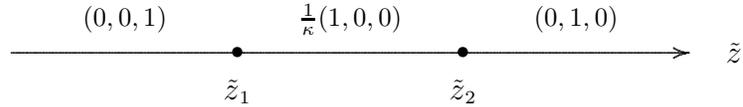}
\caption{The rod structure of the five-dimensional Schwarzschild black hole and the Ishihara--Matsuno black hole in standard orientation.}
\end{center}
\end{figure}

\newsection{Black holes on the self-dual Taub-NUT instanton}

The well-known Gross--Perry--Sorkin magnetic monopole \cite{Gross:1983,Sorkin:1983}, also known as the Kaluza--Klein monopole, when lifted to five dimensions in a regular fashion, is nothing but the five-dimensional space-time obtained by adding a flat time dimension to the self-dual Taub-NUT instanton. The simplest black hole sitting on the self-dual Taub-NUT instanton is then the magnetically charged static Kaluza--Klein black hole, lifted to five dimensions. In the form given by Ishihara and Matsuno \cite{Ishihara:2005}, it has the following metric:\footnote{Note, however, that we have rescaled the time coordinate so that $\frac{\partial}{\partial t}$ is normalised at infinity.}
\ba
\label{bh_self-dual_TN}
{\dif s}^{2}&=&- \left( 1-{\frac {r_0^{2}}{{r}^{2}}} \right)k(r_0)\, {{\dif t}}^{2}+\frac{{r}^{2}}{4} \left( {\dif \psi}+\cos \theta \,{\dif \phi}
 \right) ^{2}+\frac{{k(r)}^{2}}{ 1-{\frac {r_0^{2}}{{
r}^{2}}}}\, {{\dif r}}^{2}+\frac{k(r){r}^{2}}{4} \left( {{\dif \theta}}^{2}+
\sin ^{2} \theta\, {{\dif \phi}}^{2}
 \right),\cr
&&
\ea
where $k(r)$ is a function defined as
\be
k(r)=\frac{r_{\infty}^2 (r_{\infty}^2-r_0^2)}{(r_{\infty}^2-r^2)^2}\,.
\ee
The parameters $r_0$, $r_\infty$ and coordinates $t$, $r$, $\theta$ take the ranges $-\infty<t<\infty$, $0\leq r_0\leq r\leq r_{\infty}$, $0\leq \theta \leq \pi$. The black-hole horizon and physical infinity are located at $r=r_0$ and $r=r_{\infty}$, respectively.

The Weyl--Papapetrou coordinates $(t,\psi,\phi,\rho,z)$ are related to the above coordinates by
\ba
\label{Weyl_bh_self_TN}
\rho=\frac{r \sqrt{(r^2-r_0^2)k(r)k(r_0)}}{4}\, \sin\theta\,,\qquad
 z=\frac{2r_{\infty}^2 r^2-r_0^2 r^2-r_0^2 r_{\infty}^2}{8(r_{\infty}^2-r^2)}\sqrt{k(r_0)}\,\cos\theta\,.
\ea
In these coordinates, the rod structure has two turning points, at $(\rho=0,z=z_1\equiv-\frac{\sqrt{k(r_0)}\,r_0^2}{8})$ or $(r=r_0,\theta=\pi)$, and $(\rho=0,z=z_2\equiv\frac{\sqrt{k(r_0)}\,r_0^2}{8})$ or $(r=r_0, \theta=0)$. These two turning points correspond to the south and north poles of the black-hole horizon, respectively. From left to right, the three rods are:

\begin{itemize}
\item Rod 1: a semi-infinite space-like rod located at $(\rho=0, z\leq z_1)$ or $(r\geq r_0,\theta=\pi)$, with direction $\ell_1=(0,1,1)$.

\item Rod 2: a finite time-like rod located at $(\rho=0, z_1\leq z\leq z_2)$ or $(r=r_0, 0\leq \theta\leq \pi)$, with direction $\ell_2=\frac{1}{\kappa}(1,0,0)$, where $\kappa=\frac{1}{\sqrt{k(r_0)}\,r_0}$ is the surface gravity on the black-hole horizon represented by this rod.

\item Rod 3: a semi-infinite space-like rod located at $(\rho=0, z\geq z_2)$ or $(r\geq r_0,\theta=0)$, with direction $\ell_3=(0,-1,1)$.
\end{itemize}

To ensure regularity, we identify the orbits generated by $\{\ell_1, \ell_3\}$ with period $2\pi$ independently, although in general, the space-time could be further quotiented by a $\mathbb{Z}_s$ group for any natural number $s$, as in the case of the five-dimensional Schwarzschild black hole. Thus the following identifications on the coordinates $(\psi,\phi)$ are made to ensure regularity:
\be
(\psi, \phi) \rightarrow (\psi+4 \pi, \phi) \,,\qquad (\psi, \phi) \rightarrow (\psi+2 \pi,\phi+2 \pi) \,.
\ee
The direction-pair $\{\ell_1,\ell_3\}$ is then identified as the pair of independent $2\pi$-periodic generators of the $U(1)\times U(1)$ isometry group of the space-time.

We can put the rod structure in standard orientation by taking $\{\tilde{V}_{(0)}=\frac{\partial}{\partial t},\tilde{V}_{(1)}=\ell_3,\tilde{V}_{(2)}=\ell_1\}$. The corresponding new Weyl--Papapetrou coordinates $(\tilde{t},\tilde{\psi},\tilde{\phi},\tilde{\rho},\tilde{z})$ are related to the old coordinates (\ref{Weyl_bh_self_TN}) by
\be
t=\tilde{t}\,,\quad
\psi=-\tilde{\psi}+\tilde{\phi}\,,\quad
\phi=\tilde{\psi}+\tilde{\phi}\,,\quad
\rho=\frac{1}{2}\,\tilde{\rho}\,,\quad
z=\frac{1}{2}\,\tilde{z}\,.
\ee
The two turning points are now pushed to $(\tilde{\rho}=0,\tilde{z}=\tilde{z}_1\equiv-\frac{\sqrt{k(r_0)}\,r_0^2}{4})$ and $(\tilde{\rho}=0,\tilde{z}=\tilde{z}_2\equiv\frac{\sqrt{k(r_0)}\,r_0^2}{4})$, and the corresponding directions of the three rods from left to right are $K_1=(0,0,1)$, $K_2=\frac{1}{\kappa}(1,0,0)$ and $K_3=(0,1,0)$. This is illustrated in Fig.~1. In the new Weyl--Papapetrou coordinates, the following identifications are made to ensure regularity:
\be
\label{identification for self-dual Taub-NUT 4}
(\tilde{\psi}, \tilde{\phi}) \rightarrow (\tilde{\psi},\tilde{ \phi}+2 \pi) \,,\qquad (\tilde{\psi}, \tilde{\phi}) \rightarrow (\tilde{\psi}+2 \pi, \tilde{\phi}) \,.
\ee
We can see that this solution is fully characterised by its rod structure, i.e., given any positive value of $\kappa$ and $\tilde{z}_2$ satisfying $\kappa^2 \tilde{z}_2\leq \frac{1}{4}$, we can find only one solution corresponding to these parameters. The degeneracy of the rod structure only occurs when the black hole is absent, i.e., in the background space-time \cite{Chen:2010}. This background space-time is obtained by taking $r_0=0$, which is nothing but the direct product of a flat time dimension and the self-dual Taub-NUT instanton in a new form.

It is obvious that the black hole has horizon-topology $S^3$. At infinity $r\rightarrow r_{\infty}$, the space-time approaches the direct product of a flat time dimension and the asymptotic structure of the self-dual Taub-NUT instanton with NUT charge $n=\frac{r_{\infty}}{4}$ \cite{Ishihara:2005}. In other words, the space-time is asymptotically a non-trivial finite $S^1$ fibre bundle over $M^{1,3}$. $\frac{\partial}{\partial \psi}$ generates the $S^1$ fibre at infinity, with a constant size $2\pi r_{\infty}$. By taking the NUT charge to infinity, so that $r_{\infty}\rightarrow \infty$, $k(r_0)=k(r)=1$, we obviously recover the five-dimensional Schwarzschild black hole.

Angular momenta can be added to the solution (\ref{bh_self-dual_TN}) \cite{Wang:2006}, which, when dimensionally reduced to four dimensions, has the interpretation of a non-rotating dyonic Kaluza--Klein black hole \cite{Gibbons:1985}. In our context, its rod structure can be easily analyzed, and it can be classified as a stationary black hole on the self-dual Taub-NUT instanton with one rotational parameter. The general stationary black hole on the self-dual Taub-NUT instanton with two independent rotational parameters is then the rotating dyonic Kaluza--Klein black hole \cite{Rasheed:1995,Matos:1996,Larsen:1999}, if lifted to five dimensions with the regularity conditions imposed appropriately. By taking the NUT charge of the space-time to infinity, the five-dimensional Myers--Perry black hole is recovered \cite{Itzhaki:1998,Emparan:2006,Emparan:2007}. In these cases, the direction of the time-like rod is rotated, i.e., it has components involving $\ell_1$ and $\ell_3$, so the black hole has non-vanishing angular velocities along the two space-like directions.

The general rotating dyonic Kaluza--Klein black hole, as a regular five-dimensional space-time, is a stationary black hole sitting at the turning point of the self-dual Taub-NUT instanton \cite{Itzhaki:1998,Emparan:2006,Emparan:2007}. In terms of the rod structure, a black hole can also sit on one of the semi-infinite rods of this instanton, resulting in a black-ring solution. The static class was constructed by Ford et al.~\cite{Ford:2007}, which is singular because of the presence of a conical singularity. By adding angular momenta to balance its self-gravitation, Camps et al.\ \cite{Camps:2008} constructed a regular rotating black ring on the self-dual Taub-NUT instanton. But their solution is probably not the most general solution in this class, as it has only one rotational parameter. A more general solution with two independent rotational parameters, classified as a double-rotating black ring on the self-dual Taub-NUT instanton, is expected to exist, but has not been constructed explicitly. If such a solution exists, we would be able to recover the double-rotating Pomeransky--Sen'kov black ring \cite{Pomeransky:2006} by taking the NUT charge to infinity. Also, multi-black-hole solutions on the self-dual Taub-NUT instanton may exist. In all these configurations, by removing the black holes/rings, we recover the self-dual Taub-NUT instanton with a flat time dimension.

We emphasise that the NUT charge of the above solutions characterises the size of the compact (space-like) dimension at infinity for these space-times, which are completely regular as five-dimensional solutions. It is not of the more familiar type associated with Dirac--Misner singularities \cite{Misner:1963}, since for all the solutions classified in this section, there are no asymptotic cross terms involving time and any of the spatial dimensions. Thus if we perform Kaluza--Klein reduction along the compact dimension, we obtain four-dimensional asymptotically flat space-times, {\it without\/} NUT charges. In particular, the time dimension is non-compact, and has the range $-\infty<t<\infty$. On the other hand, there do exist five-dimensional solutions which have cross terms involving time and the other spatial dimensions. If Kaluza--Klein reduction is performed on them, we obtain four-dimensional space-times with real NUT charges in general. So in these cases, the time dimension is probably identified and is thus compact. In fact, the Kaluza--Klein black holes in \cite{Gibbons:1985,Rasheed:1995,Larsen:1999} were obtained by imposing the condition that these real NUT charges vanish.

\newsection{Black holes on the Euclidean Schwarzschild instanton}

The solution describing a static black hole sitting on the Euclidean Schwarzschild instanton was first constructed by Emparan and Reall \cite{Emparan:2001b}. It will not be written down explicitly here, as it can be obtained as a special case of the solution to be presented in the following section, describing a static black hole sitting on the Euclidean Kerr instanton. The rod structure of this solution is illustrated in Fig.~2(a).

The background space-time, consisting of the direct product of a flat time dimension and the Euclidean Schwarzschild instanton, is asymptotically $M^{1,3}\times S^1$ with a finite and constant $S^1$. The finite space-like rod (rod 3 in Fig.~2(a)) is a minimal $S^2$ surface which can be interpreted as a static ``bubble of nothing'' \cite{Witten:1981}. The above-mentioned black-hole solution of Emparan and Reall \cite{Emparan:2001b} can therefore be described as a black hole on a Kaluza--Klein bubble.

A second black hole can be added at the second turning point of the Euclidean Schwarz\-schild instanton, to obtain a static solution describing two black holes on a Kaluza--Klein bubble \cite{Elvang:2002}. Rotating generalisations were constructed in \cite{Tomizawa:2007,Iguchi:2007b}.
In terms of the rod structure, the finite rod of the Euclidean Schwarzschild instanton can also be cut and placed with black holes \cite{Elvang:2004}. Placing a single black hole would result in a black ring surrounded by two Kaluza--Klein bubbles; such a solution with one rotational parameter was recently constructed in \cite{Nedkova:2010}. By removing the black holes/rings from any of these space-times, we recover the Euclidean Schwarzschild instanton with a flat time dimension.

\begin{figure}[t]
\begin{center}
\includegraphics{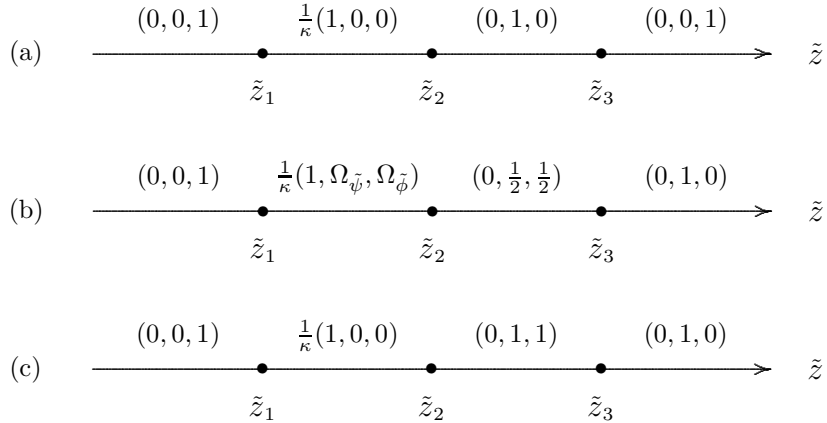}
\caption{The rod structure of: (a) a black hole on the Euclidean Schwarzschild or Kerr instanton; (b) a (rotating) black hole on the Eguchi--Hanson instanton; and (c) a black hole on the Taub-bolt instanton; all in standard orientation.}
\end{center}
\end{figure}

\newsection{Black holes on the Euclidean Kerr instanton}

The solution describing a static black hole sitting on the Euclidean Kerr instanton has the following metric in C-metric-like coordinates:\footnote{This solution can be obtained from (5.20) in Ford et al.~\cite{Ford:2007} by imposing $v_{<}=v_{>}$. However, it was not studied in any detail in \cite{Ford:2007}, being used as stepping-stone to construct a black ring on Taub-NUT space. Furthermore, it was presented in a different, rather more complicated, form than the new one used here. The C-metric-like coordinates used in (\ref{bh_EK}) are described in Appendix H of \cite{Harmark:2004}. We remark that the solution can also be obtained directly using the inverse-scattering method, by starting from a suitable seed solution.}
\ba
\label{bh_EK}
{\dif s}^{2}&=&-{\frac{1+cy}{1+cx}}\,{\dif t}^{2}+{
\frac {F( x,y ) }{H ( x,y) }} \left( {\dif\psi}+\Omega\right) ^{2}\cr
 &&+\frac{2{\varkappa}^{4} \left( 1+cx
 \right) H ( x,y )}{c^2 \left( 1-c \right) \left( 1-{\alpha}^{2}
 \right) \left( 1+{\alpha}^{2} \right) ^2\left( x-y \right)
^{3}}  \left( {\frac {{\dif x}^{2}}{G
 ( x ) }}-{\frac {{\dif y}^{2}}{G ( y ) }}+A\,{{
\dif\phi}}^{2} \right),
\ea
where $\Omega$ and $A$ are defined as
\ba
\Omega&=&{\frac {2\alpha c^2 \varkappa^{2}\, [ 1+c- \left( 1-c
 \right) {\alpha}^{2} ][ 1-c- \left( 1+c \right) {\alpha}
^{2} ] }{
  \left( 1+{\alpha}^{2}
 \right) }}\cr
 &&\times \frac{\left( 1+y \right) [  \left( 1-y \right)
 \left( 2-c+cx \right)+\left( 1-x \right)  \left( 2
-c +cy\right) {\alpha}^{2} ] G ( x )}{\left( 1-x \right)  \left( x-y \right) F( x,y ) } \,\dif\phi\,,\cr
A&=&-{\frac { 2 c^2\left( 1-c \right)  \left( 1-{\alpha}^{2} \right)
 \left( 1+{\alpha}^{2} \right) ^{2}G \left( x \right) G ( y
 ) }{ \left( x-y \right)  \left( 1+cy \right) F ( x,y
 ) }}
 \,.
\ea
The functions $G(x)$, $H(x,y)$ and $F(x,y)$ are defined as
\ba
G ( x ) &=& \left( 1+cx \right)  ( 1-{x}^{2} )\,,\cr
H ( x,y ) &=& \left( 1+cx \right)  [  \left( 1-c \right)  \left( 1+c- \left( 1-c \right) {\alpha}^{2} \right) -
 \left( 1+cy
 \right) \left( 1-c- \left( 1+c \right) {\alpha}^{2} \right)   ] ^{2}\cr
&& -{\alpha}^{2} \left( 1+cy \right)  [\left( 1-c \right)  \left( 1-c- \left( 1+c \right) {\alpha}
^{2} \right)  -
  \left( 1+cx
 \right) \left( 1+c- \left( 1-c \right) {\alpha}^{2} \right) ] ^{2},\cr
F ( x,y ) &=&{\frac {{c}^{2} \left( 1-{\alpha}^{2} \right)  \left( 1+cx \right) }{1+c}}[  \left( 1-c \right)  \left( 1-x \right)
 \left( 1-y \right)  \left( 1-c- \left( 1+c \right) {\alpha}^{2}
 \right)  \left( 1+c- \left( 1-c \right) {\alpha}^{2} \right)\cr
&& \hskip1.4in -8\,{
\alpha}^{2} \left( c+x+y+cxy \right)  ] \,.
\ea
The parameters $\varkappa$, $c$, $\alpha$ and coordinates $t$, $x$, $y$ take the ranges $\varkappa>0$, $0\leq c < 1$, $\alpha^2<\frac{1-c}{1+c}$, $-\infty<t<\infty$, $-1\leq x \leq 1$, $-\frac{1}{c}\leq y\leq -1$. The black-hole horizon is located at $y=-\frac{1}{c}$, while physical infinity is located at $(x,y)=(-1,-1)$.

The Weyl--Papapetrou coordinates $(t,\psi,\phi,\rho,z)$ are related to the above C-metric-like coordinates by
\ba
\label{C-metric}
\rho = \frac{2\varkappa^2 \sqrt{-G(x) G(y)}}{(x-y)^2}\,,\qquad
 z   =\frac{\varkappa^2 (1-xy) (2+cx+cy)}{(x-y)^2}\,.
\ea
In these coordinates, the rod structure has three turning points, at $(\rho=0,z=z_1\equiv-c\varkappa^2)$ or $(x=-1,y=-\frac{1}{c})$, $(\rho=0,z=z_2\equiv c\varkappa^2)$ or $(x=1,y=-\frac{1}{c})$,  and $(\rho=0,z=z_3\equiv \varkappa^2)$ or $(x=1,y=-1)$. They divide the $z$-axis into four rods:

\begin{itemize}
\item Rod 1: a semi-infinite space-like rod located at $(\rho=0, z\leq z_1)$ or $(x=-1,-\frac{1}{c}\leq  y< -1)$, with direction $\ell_1=(0,0,1)$.

\item Rod 2: a finite time-like rod located at $(\rho=0, z_1\leq z\leq z_2)$ or $(-1\leq x\leq 1,y=-\frac{1}{c})$, with direction $\ell_2=\frac{1}{\kappa}(1,0,0)$, where
\be
\kappa={\frac { \left(1+{\alpha}^{2} \right) \sqrt {2c \left( 1
+c \right)  \left( 1-{\alpha}^{2} \right) }}{4c{\varkappa}^{2} \left( 1+c-
 \left( 1-c \right) {\alpha}^{2} \right) }}\,,
\ee
is the surface gravity on the black-hole horizon represented by this rod.

\item Rod 3: a finite space-like rod located at $(\rho=0, z_2\leq z\leq z_3)$ or $(x=1,-\frac{1}{c}\leq y\leq -1)$, with direction $\ell_3=(0,\frac{1}{\kappa_E},\frac{\Omega_E}{\kappa_E})$, where $\kappa_E$ and $\Omega_E$ are defined as
\ba
\label{kappa_E}
{\kappa_E}&=&{\frac { \left( 1-{\alpha}^{2} \right)  \left( 1+{
\alpha}^{2} \right) ^{2}\sqrt {1-{c}^{2}}}{2{\varkappa}^{2} \left( 1-c-
 \left( 1+c \right) {\alpha}^{2} \right)  \left( 1+c- \left( 1-c
 \right) {\alpha}^{2} \right) }}\,,\cr
{\Omega_E}&=&{\frac {\alpha\, \left( 1-{\alpha}^{2} \right)  \left(
1+{\alpha}^{2} \right) }{{\varkappa}^{2} \left( 1-c- \left( 1+c \right) {
\alpha}^{2} \right)  \left( 1+c- \left( 1-c \right) {\alpha}^{2}
 \right) }}\,.
\ea

\item Rod 4: a semi-infinite space-like rod located at $(\rho=0, z\geq z_3)$ or $(-1<x\leq 1,y=-1)$, with direction $\ell_4=(0,0,1)$.
\end{itemize}

To ensure regularity, the orbits generated by the direction-pair $\{\ell_3, \ell_4\}$ of adjacent space-like rods should be identified with period $2\pi$ independently. Equivalently, the following identifications on the coordinates $(\psi,\phi)$ should be made:
\be
\label{Identifications for BH on EK}
(\psi, \phi) \rightarrow (\psi, \phi+2 \pi) \,,\qquad (\psi, \phi) \rightarrow \left(\psi+\frac{2\pi}{\kappa_E}, \phi+\frac{2\pi \Omega_E}{\kappa_E}\right) .
\ee
The direction-pair $\{\ell_3, \ell_4\}$ is then identified as the pair of independent $2\pi$-periodic generators of the $U(1)\times U(1)$ isometry group of the space-time.

By setting $\alpha=0$, we directly recover a static black hole on the Euclidean Schwarzschild instanton \cite{Emparan:2001b} from the above solution. The background space-time, obtained by setting $c=0$ so that the time-like rod vanishes, is nothing but the Euclidean Kerr instanton with a flat time dimension. The following redefinitions of parameters and coordinate transformations:
\be
m=\frac{\varkappa^2}{2}\frac{1-\alpha^2}{1+\alpha^2}\,,\qquad
a={\frac {\alpha {\varkappa}^{2}}{1+{\alpha}^{2}}}\,,\qquad
r={\frac {{\varkappa}^{2} [ 1-y+{\alpha}^{2} \left( 1-x \right)
  ]}{ \left( 1+{\alpha}^{2} \right) \left( x-y \right)  }}
\,,\qquad
\cos\theta=\frac {2+x+y}{x-y}\,,
\ee
are needed to bring the Euclidean Kerr instanton in the background space-time to the standard form:
\be
\label{Euclidean Kerr}
{\dif s}_{(4)}^{2}={\frac {{\Delta} \left( {\dif\psi}+a \sin ^{2}
 \theta\, {\dif\phi} \right) ^{2}}{{\Sigma}}}+{\frac { \sin ^{2} \theta\,
 [a{\dif\psi}- \left( {r}^{2}-{a}^{2} \right) {\dif\phi}] ^{2}}{{\Sigma}}}+{\Sigma} \left( {\frac {{\dif r}
^{2}}{{\Delta}}}+{{\dif\theta}}^{2} \right),
\ee
where $\Sigma$ and $\Delta$ are defined as
\be
{\Sigma}={r}^{2}-{a}^{2} \cos ^{2} \theta \,,\qquad {\Delta}={r}^{2}-2mr-{a}^{2}.
\ee
We thus naturally interpret the above solution as a static black hole sitting at the first turning point of the Euclidean Kerr instanton.

We can put the rod structure in standard orientation by taking $\{\tilde{V}_{(0)}=\frac{\partial}{\partial t},\tilde{V}_{(1)}=\ell_3,\tilde{V}_{(2)}=\ell_1\}$. The corresponding new Weyl--Papapetrou coordinates $(\tilde{t},\tilde{\psi},\tilde{\phi},\tilde{\rho},\tilde{z})$ are related to the old coordinates (\ref{C-metric}) by
\be
t=\tilde{t}\,,\qquad
\psi=\frac{1}{\kappa_E}\,\tilde{\psi}\,,\qquad
\phi=\frac{\Omega_E}{\kappa_E}\,\tilde{\psi}+\tilde{\phi}\,,\qquad
\rho=\kappa_E \tilde{\rho}\,,\qquad
z=\kappa_E \tilde{z}\,.
\ee
The three turning points are now pushed to $(\tilde{\rho}=0,\tilde{z}=\tilde{z}_i\equiv\frac{z_i}{\kappa_E})$ for $i=1,2,3$, and the corresponding directions of the four rods from left to right are $K_1=(0,0,1)$, $K_2=\frac{1}{\kappa}(1,0,0)$, $K_3=(0,1,0)$ and $K_4=(0,0,1)$. This is illustrated in Fig.~2(a). In the new Weyl--Papapetrou coordinates, the following identifications should be made to ensure regularity:
\be
\label{Identifications for EK 2}
(\tilde{\psi}, \tilde{\phi}) \rightarrow (\tilde{\psi},\tilde{ \phi}+2 \pi) \,,\qquad (\tilde{\psi}, \tilde{\phi}) \rightarrow (\tilde{\psi}+2 \pi, \tilde{\phi}) \,.
\ee
As in the static black hole on the self-dual Taub-NUT instanton, no degeneracy of the rod structure can be found except in the background space-time \cite{Chen:2010}.

It can be checked that this space-time is free of curvature singularities and closed time-like curves (CTCs) outside the event horizon, and is thus completely regular. It is obvious that the black hole has horizon-topology $S^3$. At infinity $(x,y)\rightarrow(-1,-1)$, the space-time approaches the direct product of a flat time dimension and the asymptotic structure of the Euclidean Kerr instanton (note that $\Omega_E$ and $\kappa_E$ now take the new values (\ref{kappa_E})) \cite{Chen:2010}. More specifically, by introducing new coordinates $\big\{r=\frac{2\varkappa^2(1-c)}{x-y},\cos\theta=\frac{2+x+y}{x-y}\big\}$, at infinity $r\rightarrow \infty$, the metric (\ref{bh_EK}) approaches
\ba
\dif s^2&\rightarrow&-\dif t^2+\dif r^2+r^2(\dif \theta^2+\sin^2\theta\,\dif \phi^2)+\dif\psi^2\cr
&\rightarrow&-\dif {\tilde{t}}^2+\dif r^2+r^2(\dif \theta^2+\sin^2\theta\,\dif \tilde{\phi}^2)+\frac{\Omega_E^2}{\kappa_E^2}\,r^2\sin^2\theta\,\dif{\tilde{\psi}}^2+\frac{2\Omega_E}{\kappa_E}\,r^2\sin^2\theta\,\dif\tilde{\psi}\dif\tilde{\phi}\,.~~~~~~
\ea
Along with the identifications (\ref{Identifications for BH on EK}) or (\ref{Identifications for EK 2}), it can be seen the space-time is asymptotically $M^{1,3}\times S^1$, with the Killing vector field $\frac{\partial}{\partial{\tilde\phi}}$ or $\frac{\partial}{\partial{\phi}}$ generating the axial symmetry of the Minkowski space-time $M^{1,3}$, and $\frac{\partial}{\partial{\tilde\psi}}$, whose orbits close off on rod 3, generating the $S^1$ that in general blows up at infinity \cite{Aharony:2002}. However, if $\frac{\Omega_E}{\kappa_E}$ is a rational number, the Killing vector field $\frac{\partial}{\partial{\psi}}$ generates closed and finite orbits (i.e., $S^1$) at infinity \cite{Hunter:1998}.

We expect that angular momenta can be added to the above black-hole solution. Also, another black hole may be added to the second turning point of the Euclidean Kerr instanton, resulting in a double-black-hole solution.

\newsection{Black holes on the Eguchi--Hanson instanton}

One may expect that, similarly as done in the previous section, black holes can be placed at the turning points of the Eguchi--Hanson instanton. However, no static black-hole solutions of such a class have been found, despite all efforts we have made. It is not clear to us whether it is because of the limitations of the solution-generating techniques we have applied, or because these configurations do not exist at all.

It turns out, however, that a {\it rotating\/} black hole on the Eguchi--Hanson instanton does exist, and its local metric is contained in the rotating-black-lens solution found in \cite{Chen:2008}. In C-metric-like coordinates, it has the form:
\ba
\label{bl_metric}
\dif s^2&=&-\frac{H(y,x)}{H(x,y)}(\dif t-\omega_\psi\,\dif\psi-\omega_\phi\,\dif\phi)^2
-\frac{F(x,y)}{H(y,x)}\,\dif\psi^2+2\frac{J(x,y)}{H(y,x)}\,\dif\psi\dif\phi\cr
&&+\frac{F(y,x)}{H(y,x)}\,\dif\phi^2+\frac{\varkappa^2H(x,y)}{2(1-a^2)(1-b)^3(x-y)^2}\left(\frac{\dif x^2}{G(x)}-\frac{\dif y^2}{G(y)}\right),
\ea
where
\ba
\omega_\psi&=&\frac{2\varkappa}{H(y,x)}\sqrt{\frac{2b(1+b)(b-c)}{(1-a^2)(1-b)}}\,
(1-c)(1+y)\{2[1-b-a^2(1+bx)]^2(1-c)\cr
&&\hskip2.1in~-a^2(1-a^2)b(1-b)(1-x)(1+cx)(1+y)\}\,,\cr
\omega_\phi&=&\frac{2\varkappa}{H(y,x)}\sqrt{\frac{2b(1+b)(b-c)}{(1-a^2)(1-b)}}\,
a(1-c)(1+x)^2(1+y)[a^4(1+b)(b-c)\cr
&&\hskip2.1in~+a^2(1-b)(-b+cb+2c)-(1-b)^2c]\,.
\ea
The functions $G(x)$, $H(x,y)$, $F(x,y)$ and $J(x,y)$ are defined as
\ba
G(x)&=&(1+cx)(1-x^2)\,,\cr
H(x,y)&=&4(1-b)(1-c)(1+bx)\{(1-b)(1-c)-a^2[(1+bx)(1+cy)+(b-c)(1+y)]\}\cr
&&+a^2(b-c)(1+x)(1+y)\{(1+b)(1+y)[(1-a^2)(1-b)c(1+x)+2a^2b(1-c)]\cr
&&-2b(1-b)(1-c)(1-x)\}\,,\cr
F(x,y)&=&\frac{2\varkappa^2}{(1-a^2)(x-y)^2}\,\Big[4(1-c)^2(1+bx)[1-b-a^2(1+bx)]^2G(y)\cr
&&~-a^2G(x)(1+y)^2\Big([1-b-a^2(1+b)]^2(1-c)^2(1+by)-(1-a^2)(1-b^2)\times\cr
&&~\times(1+cy)\{(1-a^2)(b-c)(1+y)+[1-3b-a^2(1+b)](1-c)\}\Big)\Big]\,,\cr
J(x,y)&=&\frac{4\varkappa^2a(1-c)(1+x)(1+y)}{(1-a^2)(x-y)}\,[1-b-a^2(1+b)][(1-b)c+a^2(b-c)]\times\cr
&&~\times[(1+bx)(1+cy)+(1+cx)(1+by)+(b-c)(1-xy)]\,.
\ea
The parameters $\varkappa$, $a$, $b$, $c$ and coordinates $t$, $x$, $y$ take the ranges $\varkappa>0$, $-1\leq a \leq 1$, $0\leq c \leq b<1$, $-\infty<t<\infty$, $-1\leq x \leq 1$, $-\frac{1}{c}\leq y \leq -1$. The black-hole horizon is located at $y=-\frac{1}{c}$, while physical infinity is located at $(x,y)=(-1,-1)$.

The Weyl--Papapetrou coordinates $(t,\psi,\phi,\rho,z)$ are related to the above C-metric-like coordinates by the relation (\ref{C-metric}). In these coordinates, the locations of the three turning points in the rod structure are the same as those in the static black hole on the Euclidean Kerr instanton in Sec.~5. They divide the $z$-axis into four rods:

\begin{itemize}
\item Rod 1: a semi-infinite space-like rod located at $(\rho=0, z\leq z_1)$ or $(x=-1,-\frac{1}{c}\leq  y< -1)$, with direction $\ell_1=(0,0,1)$.

\item Rod 2: a finite time-like rod located at $(\rho=0, z_1\leq z\leq z_2)$ or $(-1\leq x\leq 1,y=-\frac{1}{c})$, with direction $\ell_2=\frac{1}{\kappa}(1,\Omega_{\psi},\Omega_{\phi})$, where the surface gravity on the black-hole horizon $\kappa$ and the angular velocities $\Omega_{\psi}$, $\Omega_{\phi}$ are given by
\ba
\kappa&=&\frac{1}{4\varkappa}\sqrt{\frac{2c(1-a^2)}{b(1+b)}}\frac{(1-b)^2(1+c)}{(1-c)[(1-b)c+a^2(b-c)]}\,,\cr
\Omega_{\psi}&=&\frac{1}{\varkappa}\sqrt{\frac{(1-b)(b-c)}{2(1-a^2)b(1+b)}}\frac{1}{1-c}\,,\cr
\Omega_{\phi}&=&\frac{1}{2\varkappa}\sqrt{\frac{(1-b)(b-c)}{2(1-a^2)b(1+b)}}\frac{a[1-b-a^2(1+b)]}{(1-b)c+a^2(b-c)}\,.
\ea

\item Rod 3: a finite space-like rod located at $(\rho=0, z_2\leq z\leq z_3)$ or $(x=1,-\frac{1}{c}\leq y\leq -1)$, with direction $\ell_3=(0,\frac{n}{m},\frac{1}{m})$, where
\be
\label{bl_conditions}
n=\frac{2a[(1-b)c+a^2(b-c)]}{[1-b-a^2(1+b)](1-c)}\,,\qquad
m=\sqrt{\frac{1-b}{1+b}}\frac{(1-a^2)(1-b)(1+c)}{[1-b-a^2(1+b)](1-c)}\,.
\ee

\item Rod 4: a semi-infinite space-like rod located at $(\rho=0, z\geq z_3)$ or $(-1<x\leq 1,y=-1)$, with direction $\ell_4=(0,1,0)$.
\end{itemize}

Now, the rotating black hole on the Eguchi--Hanson instanton is obtained from (\ref{bl_metric}) when the parameters $a$ and $b$ are fixed as
\be
a=\frac{3(1-c)}{3+5c}\,,\qquad b=\frac{4c(3-c)}{5c^2-6c+9}\,,
\ee
which are determined to solve the conditions $n=1$ and $m=2$ in (\ref{bl_conditions}).\footnote{It can be easily checked that $0<a\leq \sqrt{(1-b)/(1+b)}$, so this solution is in Range I as defined in \cite{Chen:2008}.} These conditions ensure that the black hole has horizon-topology $L(1,2)\cong S^3$, and that the finite space-like rod has the correct direction $(0,\frac{1}{2},\frac{1}{2})$ required of an Eguchi--Hanson instanton background \cite{Chen:2010}. The resulting rod structure, which is already in standard orientation, is illustrated in Fig.~2(b) (with tildes added to $z$, $\psi$ and $\phi$).

To ensure regularity, the orbits generated by the direction-pair $\{\ell_3, \ell_4\}$ of adjacent space-like rods should be identified with period $2\pi$ independently. Equivalently, the following identifications on the coordinates $(\psi,\phi)$ should be made:
\be
(\psi, \phi) \rightarrow (\psi+2 \pi, \phi) \,,\qquad (\psi, \phi) \rightarrow (\psi+\pi, \phi+\pi) \,.
\ee
The direction-pair $\{\ell_3, \ell_4\}$ is then identified as the pair of independent $2\pi$-periodic generators of the $U(1)\times U(1)$ isometry group of the space-time. We note that no degeneracy of the rod structure can be found.

Outside the event horizon, this space-time is free of curvature singularities, and no CTCs have been found despite extensive numerical checks. At infinity $(x,y)\rightarrow(-1,-1)$, the space-time approaches a five-dimensional Minkowski space-time quotiented by a $\mathbb{Z}_2$ group, i.e., $M^{1,4}/\mathbb{Z}_2$. The background space-time, obtained by setting $c=0$ so that the time-like rod vanishes, is nothing but the Eguchi--Hanson instanton with a flat time dimension. The following parameter redefinition and coordinate transformations:
\be
\hat a=\varkappa\,,\qquad
r^2=\frac {\varkappa^2(2-x-y) }{ x-y}\,,\qquad
\cos\theta=\frac {2+x+y}{x-y}\,,\qquad
\hat\psi=\psi+\phi\,,\qquad
\hat\phi=-\psi+\phi\,,
\ee
are needed to bring the Eguchi--Hanson instanton in the background space-time to the standard form:
\be
{\dif s}_{(4)}^{2}=\left( 1-{\frac {\hat a^{4}}{{r}^{4}}} \right) \frac{{r}^{
2}}{4} ( {\dif\hat\psi}+\cos \theta\, {\dif\hat\phi} ) ^{2
}+\left( 1-{\frac {\hat a^{4}}{{r}^{4}}} \right) ^{-1} {\dif r}^{2}+\frac{{r}^{2}}{4} ( {{\dif\theta}}^{2}+ \sin^{2} \theta\,
 \dif\hat\phi^{2} )\,.
\ee
We thus naturally interpret the above solution as a rotating black hole sitting at the first turning point of the Eguchi--Hanson instanton.

It is instructive to see what happens if we instead begin with the static limit of the black-lens solution (\ref{bl_metric}). This limit is obtained by setting $b=c$, and its explicit form can be found in Sec.~II of \cite{Chen:2008}. To ensure that $n=1$, we have to set $a=\frac{1-c}{1+c}$, in which case $m=\frac{2}{\sqrt{1-c^2}}\geq2$. Thus, the background space-time of an Eguchi--Hanson instanton with a flat time dimension is only recovered in the limit $c=0$, when the time-like rod vanishes and there is no black hole present. In general, when there is a black hole present, the finite space-like rod does not have the correct direction $(0,\frac{1}{2},\frac{1}{2})$ required of an Eguchi--Hanson instanton background. Instead, this solution is more naturally interpreted as a black hole sitting on the possible new gravitational instanton conjectured in the Sec.~5.1 of \cite{Chen:2010}, if we set $m\equiv p\geq 3$ to be an integer so that the finite space-like rod has direction $(0,\frac{1}{p},\frac{1}{p})$. By identifying the orbits generated by $\{\ell_3,\ell_4\}$ with period $2\pi$ independently, we get a black hole with horizon-topology $L(1,p)\cong S^3$ in an asymptotically lens-space $L(p,1)$ space-time \cite{Lu:2008}.\footnote{Since $c$ is fixed in terms of $p$, the background limit of this space-time is taken as $\varkappa\rightarrow 0$. In this limit, the finite space-like rod also vanishes, and we recover a quotiented five-dimensional Minkowski space-time $M^{1,4}/\mathbb{Z}_p$ (which contains an orbifold singularity and is thus singular). This suggests that the conjectured new gravitational instanton of \cite{Chen:2010} cannot exist alone as a background space; a black hole must be present in such a space-time through some unknown mechanism. This in turn suggests that the conjectured new gravitational instanton may not exist at all.}

A more general class of rotating black holes on the Eguchi--Hanson instanton may similarly be obtained from the double-rotating black-lens solution \cite{Chen:2008}. More complicated configurations, such as black rings and multi-black-holes on the Eguchi--Hanson instanton, may also exist, but have not been found yet.

Black holes may also possibly be constructed on the double-centered (or even multi-collinearly-centered) Taub-NUT instanton \cite{Hawking:1976}, but to the best of our knowledge no such examples have been found in the literature. If such solutions exist, taking the NUT charge to infinity would result in black holes on the Eguchi--Hanson instanton.

\newsection{Black holes on the Taub-bolt instanton}

The solution describing a static black hole sitting on the Euclidean non-self-dual Taub-NUT solution has a metric in the following form in C-metric-like coordinates:\footnote{This solution can be obtained from (5.20) in Ford et al.~\cite{Ford:2007} by imposing $\tilde{Q}=\infty$, so that $v$ in (5.26) has only a non-vanishing second component. Redefinition of parameters and coordinates are then needed to bring the solution into the new form used here. It can also be obtained directly using the inverse-scattering method, by starting from the same seed solution as alluded to in Footnote 2.}
\ba
\label{bh_TN}
{\dif s}^{2}&=&-{\frac {  1+cy }{1+cx}}\,{\dif t}^{2}+{
\frac {F ( x,y ) }{H ( x,y ) }} \left( {\dif\psi}+\Omega \right) ^{2}\cr
 &&+\frac{2{\varkappa}^{4} \left( 1-c
 \right)  \left( 1+cx \right) H ( x,y ) }{ \left( 1-{\alpha}^{2} \right) \left( x-y \right) ^3} \left( {\frac {{\dif x}^{2}}{G ( x ) }}-{\frac {{\dif y}^{2}}{G ( y
 ) }}+A\,{{\dif\phi}}^{2} \right),
\ea
where $\Omega$ and $A$ are defined as
\ba
\Omega&=&{\frac {2 \alpha\,{\varkappa}^{2} [2+x+y+c \left( 1+x
 \right)  \left( 1+y \right) ] }{ \left( 1-{\alpha}^{2}
 \right)  \left( x-y \right) }} \,\dif\phi \,,\cr
A&=&-{\frac {2 \left( 1+x \right)  \left( 1+y \right) }{ \left( 1-c
 \right)  \left( x-y \right) }}\,.
\ea
The functions $G(x)$, $H(x,y)$ and $F(x,y)$ are defined as
\ba
G ( x ) &=& \left( 1+cx \right)  ( 1-{x}^{2} )\,,\cr
H ( x,y ) &=& \left( 1+cx \right)  \left( 1-y \right)^{2}-{
\alpha}^{2} \left( 1+cy \right)  \left( 1-x \right)^{2},\cr
F ( x,y ) &=& ( 1-{\alpha}^{2} )  \left( 1-x
 \right)  \left( 1-y \right)  \left( 1+cx \right).
\ea
The parameters $\varkappa$, $c$, $\alpha$ and coordinates $t$, $x$, $y$ take the ranges $\varkappa>0$, $0\leq c < 1$, $\alpha^2<1$, $-\infty<t<\infty$, $-1\leq x \leq 1$, $-\frac{1}{c}\leq y\leq -1$. The black-hole horizon is located at $y=-\frac{1}{c}$, while physical infinity is located at $(x,y)=(-1,-1)$.

The Weyl--Papapetrou coordinates $(t,\psi,\phi,\rho,z)$ are related to the above C-metric-like coordinates by the relation (\ref{C-metric}). In these coordinates, the locations of the three turning points in the rod structure are the same as those in the static black hole on the Euclidean Kerr instanton in Sec.~5. They divide the $z$-axis into four rods:

\begin{itemize}
\item Rod 1: a semi-infinite space-like rod located at $(\rho=0, z\leq z_1)$ or $(x=-1,-\frac{1}{c}\leq  y< -1)$, with direction $\ell_1=(0,{\frac {2 \alpha{\varkappa}^{2}}{1-{\alpha}^{2}}},1)$.

\item Rod 2: a finite time-like rod located at $(\rho=0, z_1\leq z\leq z_2)$ or $(-1\leq x\leq 1,y=-\frac{1}{c})$, with direction $\ell_2=\frac{1}{\kappa_\mathrm{\it TN}}(1,0,0)$, where
\be
\kappa_\mathrm{\it TN}=\frac{1}{2\varkappa^2}\sqrt{\frac{1-\alpha^2}{2c(1+c)}}\,,
\ee
is the surface gravity on the black-hole horizon represented by this rod.

\item Rod 3: a finite space-like rod located at $(\rho=0, z_2\leq z\leq z_3)$ or $(x=1,-\frac{1}{c}\leq y\leq -1)$, with direction $\ell_3=(0,{\frac {2{\varkappa}^{2}\sqrt {1-{c}^{2}}}{1-{\alpha}^{2}}},0)$.

\item Rod 4: a semi-infinite space-like rod located at $(\rho=0, z\geq z_3)$ or $(-1<x\leq 1,y=-1)$, with direction $\ell_4=(0,-{\frac {2 \alpha{\varkappa}^{2}}{1-{\alpha}^{2}}},1)$.
\end{itemize}

By setting $\alpha=0$, we directly recover a static black hole on the Euclidean Schwarzschild instanton from the above solution. The background space-time, obtained by setting $c=0$ so that the time-like rod vanishes, is nothing but the non-self-dual Taub-NUT solution with a flat time dimension. The following redefinitions of parameters and coordinate transformations:
\be
m=\frac{\varkappa^2}{2}\frac{1+\alpha^2}{1-\alpha^2}\,,\qquad
n={\frac {{\alpha \varkappa}^{2}}{1-{\alpha}^{2}}}\,,\qquad
r={\frac {{\varkappa}^{2} [1-y-{\alpha}^{2} \left( 1-x \right) ] }{ \left( 1-{\alpha}^{2} \right)  \left( x-y \right) }}\,,\qquad
\cos\theta=\frac {2+x+y}{x-y}\,,
\ee
are needed to bring the non-self-dual Taub-NUT solution in the background space-time to the standard form:
\be
\label{non-self-dual Taub-NUT}
{\dif s}_{(4)}^{2}=f(r)  \left( {\dif\psi}+2n\cos
 \theta\, {\dif\phi} \right) ^{2}+{\frac {{\dif r}^{2
}}{f(r) }}+ \left( {r}^{2}-{n}^{2} \right)  ( {{
\dif\theta}}^{2}+\sin ^{2} \theta \,{{
\dif\phi}}^{2} )\,,
\ee
where the function $f(r)$ is defined as
\be
f(r) ={\frac {{r}^{2}+{n}^{2}-2mr}{{r}^{2}-{n}^{2}}}\,.
\ee

Similarly as in the case of the non-self-dual Taub-NUT solution, the solution (\ref{bh_TN}), in general, does not describe a regular space-time. A regular class can be obtained by fixing $\alpha=\pm \frac{\sqrt {1-{c}^{2}}}{2}$, to which we pay attention from now on. In this case, without loss of generality, the directions become $\ell_1=(0,2n,1)$, $\ell_2=\frac{1}{\kappa_\mathrm{\it TB}}(1,0,0)$, $\ell_3=(0,4n,0)$ and $\ell_4=(0,-2n,1)$, with NUT charge $n=\pm\frac{2\varkappa^2\sqrt{1-c^2}}{3+c^2}$, and surface gravity on the horizon $\kappa_\mathrm{\it TB}=\frac{1}{4\varkappa^2}\sqrt{\frac{3+c^2}{2c(1+c)}}$, so that $\ell_3=\ell_1-\ell_4$ as required of a Taub-bolt instanton background \cite{Chen:2010}. To ensure regularity, the orbits generated by the direction-pair $\{\ell_3, \ell_4\}$ of adjacent space-like rods should be identified with period $2\pi$ independently. Equivalently, the following identifications on the coordinates $(\psi,\phi)$ should be made:
\be
(\psi, \phi) \rightarrow (\psi+4 n \pi, \phi+2 \pi)\,,\qquad
(\psi, \phi) \rightarrow (\psi+8 n \pi, \phi)  \,.
\ee
The direction-pair $\{\ell_3, \ell_4\}$ is then identified as the pair of independent $2\pi$-periodic generators of the $U(1)\times U(1)$ isometry group of the space-time. The background space-time of this regular class of solutions is nothing but the Taub-bolt instanton with a flat time dimension. We thus naturally interpret this solution as a static black hole sitting at the first turning point of the Taub-bolt instanton.

We can put the rod structure in standard orientation by taking $\{\tilde{V}_{(0)}=\frac{\partial}{\partial t},\tilde{V}_{(1)}=-\ell_4,\tilde{V}_{(2)}=\ell_1\}$. The corresponding new Weyl--Papapetrou coordinates $(\tilde{t},\tilde{\psi},\tilde{\phi},\tilde{\rho},\tilde{z})$ are related to the old coordinates (\ref{C-metric}) by
\be
t=\tilde{t}\,,\qquad
\psi=2n(\tilde{\psi}+\tilde{\phi})\,,\qquad
\phi=-\tilde{\psi}+\tilde{\phi}\,,\qquad
\rho=\frac{1}{4|n|}\,\tilde{\rho}\,,\qquad
z=\frac{1}{4|n|}\,\tilde{z}\,.
\ee
The three turning points are now pushed to $(\tilde{\rho}=0,\tilde{z}=\tilde{z}_i\equiv4|n| z_i)$ for $i=1,2,3$, and the corresponding directions of the four rods from left to right are $K_1=(0,0,1)$, $K_2=\frac{1}{\kappa_\mathrm{\it TB}}(1,0,0)$, $K_3=(0,1,1)$ and $K_4=(0,1,0)$. This is illustrated in Fig.~2(c). In the new Weyl--Papapetrou coordinates, the following identifications should be made to ensure regularity:
\be
\label{identification for TB 2}
(\tilde{\psi}, \tilde{\phi}) \rightarrow (\tilde{\psi}, \tilde{\phi}+2 \pi) \,,\qquad (\tilde{\psi}, \tilde{\phi}) \rightarrow (\tilde{\psi}+2\pi,\tilde{ \phi}) \,.
\ee
Again, no degeneracy of the rod structure can be found.

It can be checked that this space-time is free of curvature singularities and CTCs outside the event horizon, and is thus completely regular. The black hole has horizon-topology $S^3$. At infinity $(x,y)\rightarrow(-1,-1)$, the space-time approaches the direct product of a flat time dimension and the asymptotic structure of the Taub-bolt instanton with NUT charge $n$ \cite{Chen:2010}. In other words, the space-time is asymptotically a non-trivial finite $S^1$ fibre bundle over $M^{1,3}$, with the $S^1$ fibre generated by the Killing vector field $\frac{\partial}{\partial \psi}$.

We expect that angular momenta can be added to the above black-hole solution. Also, another black hole may be added to the second turning point of the Taub-bolt instanton, resulting in a double-black-hole solution.

\newsection{Discussion}

In this paper, we have classified and constructed black holes on gravitational instantons with $U(1)\times U(1)$ isometry and up to two turning points.\footnote{The only exception is the double-centered Taub-NUT instanton. It would be interesting to try to find at least one example of a black-hole solution on this gravitational instanton.} These black-hole space-times possess a $\mathbb{R}\times U(1)\times U(1)$ isometry group, and are completely regular outside the event horizon (although curvature singularities are present inside the horizon). Most of the known exact five-dimensional black holes in the literature have been classified within our scheme, and three new classes of space-times were also constructed, namely black holes on the Euclidean Kerr, Eguchi--Hanson and Taub-bolt instantons, respectively.

It is known that five-dimensional black-hole space-times that asymptotically approach $M^{1,4}$ or $M^{1,3}\times S^1$ have a very rich phase structure \cite{Emparan:2008,Obers:2008,Rodriguez:2010}. The black holes/rings that have been classified or constructed in this paper have various more general asymptotic geometries. In such space-times, we can also expect a very rich black-hole phase structure. A possible extension of the present work is to then construct the most general double-rotating black hole/ring, or their superpositions, on each gravitational instanton. In particular, it would be of interest to find the explicit forms of the double-rotating black ring on Taub-NUT space, and the double-rotating black hole on the Euclidean Kerr and Taub-bolt instantons. If found, these solutions may shed light on the phase structure of space-times with the associated asymptotic geometries.
It would also be worthwhile to study the gravitational thermodynamics of such space-times. Appropriate physical quantities of these space-times, such as asymptotic mass and angular momentum, could be identified and calculated, which might then be used to characterise these solutions and obtain results concerning their uniqueness or non-uniqueness.

We note that black holes placed at a turning point (thus covering it) of the rod structure of a gravitational instanton will have horizon-topology $S^3$; while those placed somewhere along a space-like rod (thus covering part of it) will have horizon-topology $S^1\times S^2$. Thus the classification of black holes in five dimensions with $\mathbb{R}\times U(1)\times U(1)$ isometry as black holes/rings on gravitational instantons is a practical and useful scheme, albeit not a complete one. For example, if a completely regular black lens \cite{Hollands:2007,Evslin:2008,Chen:2008} does exist, it cannot be classified as a black hole on a gravitational instanton in general, since its event horizon has the topology of a lens-space $L(n,1)$, which is a $\mathbb{Z}_n$ quotient of $S^3$.

It would be interesting to seek possible interpretations in Kaluza--Klein theory of the black-hole solutions on AF/ALF gravitational instantons presented in this paper. Recall that if a flat time dimension is added to an AF/ALF gravitational instanton such as the self-dual Taub-NUT or Euclidean Kerr instanton, the resulting solution has a description in Kaluza--Klein theory in terms of a magnetic monopole or dipole, respectively \cite{Gross:1983,Sorkin:1983}. We therefore expect that the solutions describing black holes on AF/ALF gravitational instantons may be interpreted in Kaluza--Klein theory as magnetically and/or electrically charged black holes, possibly in superposition with monopoles or dipoles. We note, however, that when a five-dimensional space-time with $\mathbb{R}\times U(1)\times U(1)$ isometry is dimensionally reduced along the direction of say the Killing vector field $\frac{\partial}{\partial \psi}$, the resulting system in Kaluza--Klein theory will consist of a stationary space-time with $\mathbb{R}\times U(1)$ isometry (parameterised by $t$ and $\phi$), a gauge field and a scalar field. This means that the identifications that are made in this paper for the five-dimensional space-time to be regular, are not necessarily the same as in Kaluza--Klein theory; although in certain cases, these identifications may play a relevant role in the latter theory. So it is possible that even if we start from a completely regular five-dimensional space-time, singular objects might appear in the dimensionally reduced system in Kaluza--Klein theory.

In Sec.~4, it was mentioned that the Euclidean Schwarzschild instanton with a flat time dimension has an interpretation as a static bubble of nothing \cite{Witten:1981,Emparan:2001b}, the bubble being the minimal $S^2$ surface represented by the finite space-like rod in the rod structure. Note that a finite space-like rod (known as a ``bolt'' in the terminology of \cite{Gibbons:1979c}) also appears in the rod structures of the Euclidean Kerr, Eguchi--Hanson and Taub-bolt instantons. When a flat time dimension is added to each of these gravitational instantons, the resulting space-time can be interpreted as a static bubble of nothing, albeit with different asymptotic structures from the bubble in Euclidean Schwarzschild space.\footnote{A bubble in the Euclidean Kerr space was previously considered in \cite{Aharony:2002}, but this is an {\it expanding\/} bubble.} A black hole added to these space-times can therefore be interpreted as a black hole on the bubble. It may be worthwhile to study the bubble interpretation of these space-times in more detail.

It should also be possible to generalise the new space-times obtained in this paper to a theory including charge, such as five-dimensional Einstein--Maxwell theory or minimal supergravity. We leave these open questions for the future.

\bigskip\bigskip

{\renewcommand{\Large}{\normalsize}
}

\end{document}